\documentclass[twocolumn, times]{aastex631}
\usepackage{xspace}
\usepackage{amsmath}
\usepackage{amssymb}
\usepackage{bm}
\usepackage{hyperref}

\newcommand{\changes}[1]{\textcolor{black}{#1}}

\received{June 24, 2024}
\revised{October 12, 2024}
\accepted{November 1, 2024}
\submitjournal{AJ}

\shorttitle{Warm Super Jupiter TOI-2145b}
\shortauthors{Dong et al.}

\graphicspath{{./}{figures/}}
\begin{document}
\newcommand{\edits}[1]{\textcolor{black}{#1}}

\newcommand{\vdag}{(v)^\dagger}
\newcommand\aastex{AAS\TeX}
\newcommand\latex{La\TeX}

% Affiliations
\newcommand{\FlatironCCA}{Center for Computational Astrophysics, Flatiron Institute, 162 Fifth Avenue, New York, NY 10010, USA}
\newcommand{\UIUC}{Department of Astronomy, University of Illinois at Urbana-Champaign, Urbana, IL 61801, USA}
\newcommand{\USQ}{University of Southern Queensland, Centre for Astrophysics, West Street, Toowoomba, QLD 4350 Australia}
\newcommand{\Princeton}{Department of Astrophysical Sciences, Princeton University, 4 Ivy Lane, Princeton, NJ 08540, USA}
\newcommand{\PSUAA}{Department of Astronomy \& Astrophysics, 525 Davey Laboratory, Penn State, University Park, PA, 16802, USA}
\newcommand{\PSUCEHW}{Center for Exoplanets and Habitable Worlds, 525 Davey Laboratory, Penn State, University Park, PA, 16802, USA}
\newcommand{\PSETI}{Penn State Extraterrestrial Intelligence Center, 525 Davey Laboratory, Penn State, University Park, PA, 16802, USA}
\newcommand{\UA}{Steward Observatory, University of Arizona, 933 N.\ Cherry Ave, Tucson, AZ 85721, USA}
\newcommand{\UAA}{Department of Astronomy and Steward Observatory, University of Arizona, Tucson, AZ 85721, USA}
\newcommand{\Penn}{Department of Physics and Astronomy, University of Pennsylvania, 209 S 33rd St, Philadelphia, PA 19104, USA}
\newcommand{\Caltech}{Department of Astronomy, California Institute of Technology, Pasadena, CA 91125, USA}
\newcommand{\STScI}{Space Telescope Science Institute, 3700 San Martin Dr, Baltimore, MD 21218, USA}
\newcommand{\JHU}{Department of Physics and Astronomy, Johns Hopkins University, 3400 N Charles St, Baltimore, MD 21218, USA}
\newcommand{\GoddardESAL}{Exoplanets and Stellar Astrophysics Laboratory, NASA Goddard Space Flight Center, Greenbelt, MD 20771, USA}
\newcommand{\GoddardISTD}{Instrument Systems and Technology Division, NASA Goddard Space Flight Center, Greenbelt, MD 20771, USA}
\newcommand{\GSFC}{NASA Goddard Space Flight Center, Greenbelt, MD 20771, USA}
\newcommand{\NOIRLab}{U.S. National Science Foundation National Optical-Infrared Astronomy Research Laboratory, 950 N.\ Cherry Ave., Tucson, AZ 85719, USA}
\newcommand{\MacquarieSchool}{School of Mathematical and Physical Sciences, Macquarie University, Balaclava Road, North Ryde, NSW 2109, Australia}
\newcommand{\MacquarieCentre}{The Macquarie University Astrophysics and Space Technologies Research Centre, Macquarie University, Balaclava Road, North Ryde, NSW 2109, Australia}
\newcommand{\NIST}{National Institute of Standards \& Technology, 325 Broadway, Boulder, CO 80305, USA}
\newcommand{\CUBoulder}{Department of Physics, 390 UCB, University of Colorado, Boulder, CO 80309, USA}
\newcommand{\JPL}{Jet Propulsion Laboratory, California Institute of Technology, 4800 Oak Grove Drive, Pasadena, California 91109}
\newcommand{\MIT}{Kavli Institute for Astrophysics and Space Research, Massachusetts Institute of Technology, Cambridge, MA, USA}
\newcommand{\UCI}{Department of Physics \& Astronomy, The University of California, Irvine, Irvine, CA 92697, USA}
\newcommand{\Carleton}{Carleton College, One North College St., Northfield, MN 55057, USA}
\newcommand{\Carnegie}{Carnegie Science Earth and Planets Laboratory, 5241 Broad Branch Road, NW, Washington, DC 20015, USA}
\newcommand{\PSUICS}{Institute for Computational and Data Sciences, Penn State, University Park, PA, 16802, USA}
\newcommand{\PSUCASt}{Center for Astrostatistics, 525 Davey Laboratory, Penn State, University Park, PA, 16802, USA}
\newcommand{\NESSF}{NASA Earth and Space Science Fellow}
\newcommand{\IAS}{Institute for Advance Study, 1 Einstein Drive, Princeton, NJ 08540, USA}
\newcommand{\Tsinghua}{Department of Astronomy, Tsinghua University, Beijing 100084, China}
\newcommand{\ETH}{ETH Zurich, Institute for Particle Physics \& Astrophysics, Zurich, Switzerland}
\newcommand{\TIFR}{Department of Astronomy and Astrophysics, Tata Institute of Fundamental Research, Homi Bhabha Road, Colaba, Mumbai 400005, India}
\newcommand{\UCLA}{Department of Physics \& Astronomy, University of California Los Angeles, Los Angeles, CA 90095, USA}
\newcommand{\UAm}{Anton Pannekoek Institute for Astronomy, 904 Science Park, University of Amsterdam, Amsterdam, 1098 XH}
\newcommand{\Indiana}{Department of Astronomy, Indiana University, Bloomington, IN 47405, USA}

% Abbreviations
\newcommand{\tess}{\emph{TESS}\xspace}
\newcommand{\gaia}{\emph{Gaia}\xspace}
\newcommand{\hippa}{\emph{Hipparcos}\xspace}
\newcommand{\tres}{TRES\xspace}
\newcommand{\hires}{HIRES\xspace}
\newcommand{\neid}{NEID\xspace}
\newcommand{\ut}{UT\xspace}
\newcommand{\bjd}{BJD\xspace}
\newcommand{\filterzs}{\ensuremath{z_\mathrm{s}}}
\newcommand{\neiddrp}{$\mathtt{NEID}$-$\mathtt{DRP}$\xspace}
\newcommand{\serval}{$\mathtt{SERVAL}$\xspace}

\newcommand{\tic}{TIC-88992642\xspace}
\newcommand{\ticb}{TIC-88992642b\xspace}
\newcommand{\toi}{TOI-2145\xspace}
\newcommand{\toib}{TOI-2145b\xspace}

\providecommand{\msun}{\ensuremath{M_{\odot}}}
\providecommand{\rsun}{\ensuremath{R_{\odot}}}
\providecommand{\lsun}{\ensuremath{L_{\odot}}}
\providecommand{\rj}{\ensuremath{R_{\rm Jup}}}
\providecommand{\mj}{\ensuremath{M_{\rm Jup}}}

% Stellar parameters
% \newcommand{\ra}{$263.7581\pm0.0106$}
% \newcommand{\dec}{$+40.6950\pm0.0122$}
\newcommand{\ra}{17:35:01.94}
\newcommand{\dec}{+40:41:42.15}
\newcommand{\parallax}{$4.420\pm0.013$}

\newcommand{\mstar}{$1.71\pm0.04$}
\newcommand{\rstar}{$2.75^{+0.06}_{-0.05}$}
\newcommand{\rhostar}{$0.081\pm0.005$}
\newcommand{\logg}{$3.79\pm0.02$}
\newcommand{\teff}{$6206^{+81}_{-75}$}
\newcommand{\feh}{$+0.28^{+0.06}_{-0.05}$}
\newcommand{\age}{$1.6^{+0.2}_{-0.1}$}
\newcommand{\vsini}{$17.8\pm1.0$}

\newcommand{\Gmag}{$8.9453$}
\newcommand{\GBPmag}{$9.2178$}
\newcommand{\GRPmag}{$8.5075$}
\newcommand{\Bmag}{$9.62$}
\newcommand{\Vmag}{$9.07$}
\newcommand{\Jmag}{$8.021$}
\newcommand{\Hmag}{$7.810$}
\newcommand{\Kmag}{$7.761$}

\newcommand{\Gvbroad}{$16.53\pm1.13$}

% Planetary parameters
\newcommand{\DThiresjitter}{$23.2^{+2.3}_{-2.8}$}
\newcommand{\DTlam}{$6.8^{+2.9}_{-3.8}$}
\newcommand{\DTomega}{$96.2^{+2.4}_{-2.5}$}
\newcommand{\DTinc}{$88.6^{+0.7}_{-0.6}$}
\newcommand{\DTmp}{$5.68^{+0.37}_{-0.34}$}
\newcommand{\DTvsini}{$18.06^{+0.36}_{-0.39}$}
\newcommand{\DTaor}{$8.74^{+0.16}_{-0.14}$}
\newcommand{\DTecc}{$0.214^{+0.014}_{-0.014}$}
\newcommand{\DTrp}{$1.092^{+0.030}_{-0.028}$}
\newcommand{\DTb}{$0.165^{+0.069}_{-0.082}$}
\newcommand{\DTap}{$0.1117^{+0.0035}_{-0.0034}$}
\newcommand{\DTrprs}{$0.04082^{+0.00024}_{-0.00027}$}
\newcommand{\DTmidt}{$1982.49662^{+0.00055}_{-0.00054}$}
\newcommand{\DTper}{$10.261128^{+0.000009}_{-0.000007}$}

\newcommand{\DRPhiresjitter}{$23.3^{+2.5}_{-2.7}$}
\newcommand{\DRPlam}{$9.0^{+15.6}_{-13.4}$}
\newcommand{\DRPomega}{$96.0^{+2.4}_{-2.3}$}
\newcommand{\DRPinc}{$88.4^{+1.0}_{-1.0}$}
\newcommand{\DRPmp}{$5.52^{+0.35}_{-0.34}$}
\newcommand{\DRPvsini}{$18.39^{+0.94}_{-0.88}$}
\newcommand{\DRPaor}{$8.60^{+0.17}_{-0.15}$}
\newcommand{\DRPecc}{$0.224^{+0.013}_{-0.013}$}
\newcommand{\DRPrp}{$1.098^{+0.026}_{-0.028}$}
\newcommand{\DRPb}{$0.192^{+0.124}_{-0.106}$}
\newcommand{\DRPap}{$0.1098^{+0.0035}_{-0.0031}$}
\newcommand{\DRPrprs}{$0.04101^{+0.00029}_{-0.00029}$}
\newcommand{\DRPmidt}{$1982.49656^{+0.00052}_{-0.00054}$}
\newcommand{\DRPper}{$10.261131^{+0.000008}_{-0.000008}$}

\newcommand{\SERVALhiresjitter}{$23.5^{+2.4}_{-2.7}$}
\newcommand{\SERVALlam}{$10.9^{+11.8}_{-11.5}$}
\newcommand{\SERVALomega}{$95.9^{+2.4}_{-2.4}$}
\newcommand{\SERVALinc}{$88.6^{+1.1}_{-0.7}$}
\newcommand{\SERVALmp}{$5.51^{+0.31}_{-0.35}$}
\newcommand{\SERVALvsini}{$18.61^{+0.78}_{-0.95}$}
\newcommand{\SERVALaor}{$8.58^{+0.14}_{-0.13}$}
\newcommand{\SERVALecc}{$0.230^{+0.011}_{-0.012}$}
\newcommand{\SERVALrp}{$1.097^{+0.028}_{-0.026}$}
\newcommand{\SERVALb}{$0.168^{+0.108}_{-0.104}$}
\newcommand{\SERVALap}{$0.1095^{+0.0030}_{-0.0033}$}
\newcommand{\SERVALrprs}{$0.04099^{+0.00026}_{-0.00029}$}
\newcommand{\SERVALmidt}{$1982.49655^{+0.00053}_{-0.00055}$}
\newcommand{\SERVALper}{$10.261132^{+0.000008}_{-0.000008}$}

\newcommand{\neidjitter}{$8.4^{+2.3}_{-2.7}$}
\newcommand{\neidjitterr}{$6.1^{+1.6}_{-2.1}$}
\newcommand{\servaljitter}{$5.9^{+1.6}_{-2.0}$}
\newcommand{\servaljitterr}{$4.8^{+1.3}_{-1.3}$}
\newcommand{\nonrotv}{$2.64^{+0.20}_{-0.23}$}
\newcommand{\nonrotvv}{$2.34^{+0.20}_{-0.21}$}

% Title
\title{Origins of Super Jupiters: TOI-2145b Has a Moderately Eccentric and Nearly Aligned Orbit}

\correspondingauthor{Jiayin Dong}
\email{jdong@flatironinstitute.org}

% Author list
\author[0000-0002-3610-6953]{Jiayin Dong}
\altaffiliation{Flatiron Research Fellow}
\affiliation{\FlatironCCA}
\affiliation{\UIUC}

\author[0000-0003-1125-2564]{Ashley Chontos}
\altaffiliation{Henry Norris Russell Fellow}
\affiliation{\Princeton}

\author[0000-0002-4891-3517]{George Zhou}
\affiliation{\USQ}

\author[0000-0001-7409-5688]{Gudmundur Stefansson}
\affil{\UAm}

\author[0000-0002-7846-6981]{Songhu Wang}
\affiliation{\Indiana}

\author[0000-0003-0918-7484]{Chelsea X. Huang}
\affil{\USQ}

% NEID team members alphabetical order
\author[0000-0002-5463-9980]{Arvind F.\ Gupta}
\altaffiliation{NOIRLab Postdoctoral Fellow}
\affil{\NOIRLab}

\author[0000-0003-1312-9391]{Samuel Halverson}
\affil{\JPL}

\author[0000-0001-8401-4300]{Shubham Kanodia}
\altaffiliation{Carnegie EPL Fellow}
\affil{\Carnegie}

\author[0000-0002-4927-9925]{Jacob K. Luhn}
\altaffiliation{NASA Postdoctoral Program Fellow}
\affil{\UCI}
\affil{\JPL}

\author[0000-0001-9596-7983]{Suvrath Mahadevan}
\altaffiliation{NEID Principal Investigator}
\affil{\PSUAA}
\affil{\PSUCEHW}

\author[0000-0002-0048-2586]{Andrew Monson}
\affil{\UA}

\author[0000-0003-0353-9741]{Jaime A. Alvarado-Montes}
\affil{\MacquarieSchool}
\affil{\MacquarieCentre}

\author[0000-0001-8720-5612]{Joe P.\ Ninan}
\affil{\TIFR}

\author[0000-0003-0149-9678]{Paul Robertson}
\altaffiliation{NEID Instrument Team Project Scientist}
\affil{\UCI}

\author[0000-0001-8127-5775]{Arpita Roy}
\affiliation{Astrophysics \& Space Institute, Schmidt Sciences, New York, NY 10011, USA}

\author[0000-0002-4046-987X]{Christian Schwab}
\affil{\MacquarieSchool}
\affil{\MacquarieCentre}

\author[0000-0001-6160-5888]{Jason T.\ Wright}
\affil{\PSUAA}
\affil{\PSUCEHW}
\affil{\PSETI}

\begin{abstract}
Super Jupiters are giant planets with \changes{several Jupiter masses}. 
It remains an open question whether these planets originate with such high masses or grow through collisions. Previous work demonstrates that warm super Jupiters tend to have more eccentric orbits compared to regular-mass warm Jupiters. This correlation between mass and eccentricity may indicate that planet-planet interactions significantly influence the warm giant planet demographics.
\changes{Here} we conducted a detailed characterization of a warm super Jupiter, \toib. This analysis utilized previous observations from TESS and Keck/HIRES, enhanced by new Rossiter-McLaughlin effect data from the NEID spectrometer on the 3.5\,m WIYN Telescope. \toib is a \DTmp\,\mj\, planet on a moderate eccentricity ($e=$ \DTecc), 10.26-day orbit, orbiting an evolved A-star. We constrain the projected stellar obliquity to be $\lambda=$ \DTlam$\degr$ from two NEID observations.
\changes{Our $N$-body simulations suggest that the formation of super Jupiter \toib could involve either of two scenarios: a high initial mass or growth via collisions. On a population level, however, the collision scenario can better describe the mass-eccentricity distribution of observed warm Jupiters.}
\end{abstract}


\section{Introduction} \label{sec:intro}

It has long been recognized that a positive correlation between planetary mass and orbital eccentricity exists among radial velocity discovered giant planets \citep{Butler06, Wright09}. These giant planets have orbital periods ranging from a few days to several thousand days and projected mass ($M_p\sin{i}$) from roughly 0.1 to 10 \mj. This positive mass-eccentricity correlation has been interpreted as a result of planet-planet interactions, such as scatterings and collisions \citep[e.g.,][]{Ford08, Chatterjee08, Juric08, Frelikh19}.

Recently, a similar trend has been reported in the population of transiting warm Jupiters \citep{Gupta24}. Close-in giant planets with masses ranging from 0.3 to 15\,\mj\, and orbital periods between 10 and 365 days exhibit a mass-dependent eccentricity distribution. \changes{Unlike many planets discovered via radial velocity}, these transiting giant planets do not suffer from the mass degeneracy due to the unknown orbital inclination angle.
Warm Jupiters less massive than 2\,\mj\, tend to have circular or low eccentricity orbits, while those more massive than 2\,\mj-- i.e., super-Jupiters--exhibit a broad range of eccentricities. This mass-eccentricity dependence likely explains the bimodal eccentricity distribution observed in warm Jupiters \citep{Dong21}, where the observed low-$e$ component represents the low-mass warm Jupiters, while the high-$e$ component represents the super, warm Jupiters.

Planet-planet interactions likely play a role in shaping the mass and eccentricity distribution of warm Jupiters, shedding light on their origins. Among these, the formation of super Jupiters is particularly interesting. These massive planets can either form through collisions between multiple lower-mass giant planets, resulting in low eccentricity and mutual inclinations, or they may be born massive, with their eccentricity and inclination further excited by companions.
It is also unclear whether the origin of super Jupiters depends on stellar properties.
To better understand this feature, we conduct a detailed characterization of a warm, super Jupiter, \toib.
The planet was first discovered and had its orbital properties confirmed by \cite{rodriguez23}, and later had its properties refined by \cite{chontos24}.
\toib is a 10.3-day period, 5.7 Jupiter-mass planet orbiting a retired A-star ($M_\star = 1.71 \pm 0.04$ M$_\sun$, $\log{g} = 3.79\pm 0.02$). The planet has a moderate orbital eccentricity of 0.22 but unknown stellar obliquity.

The underlying assumption of planet-planet interactions as the cause of the observed mass-eccentricity trend is that dynamical interactions primarily occur at the semimajor axes of the planets observed today. Whether giant planets migrated inward or formed in situ, post-formation dynamical interactions shape the observed trend \citep{Wu23}. Under such assumptions, planet-planet interactions could excite mutual inclinations between planets, but not significantly so \citep[$i_{\rm mutual} < 40\degr$;][]{Anderson20}. This is consistent with the trend of low stellar obliquity observed in the warm Jupiter population around single stars \citep[e.g.,][]{Rice22, Dong22, Radzom24, WXY2024}.
Some warm Jupiters, such as TOI-1859b \citep{Dong23}, are found in misaligned orbits; however, their host stars often have distant stellar companions, with projected distances around 2400 au in this case. The impact of stellar companions on planet formation remains unclear.

In this work, we present the Rossiter-McLaughlin (RM) effect measurements of \toib (HIP 86040) using the high-resolution NEID spectrograph. In Section~\ref{sec:observation}, we summarize previous \tess and \hires observations. In Sections~\ref*{sec:stellar} and \ref*{sec:planet}, we model and present the stellar and planetary properties of \toib, respectively, combining the \tess transits, \hires radial velocity, and \neid RM-effect and Doppler Tomography signals. We also search for external companions of \toib using \gaia and \hippa astrometry. Lastly, in Section~\ref{sec:discussion}, we discuss the properties of the \toi system and its implications for warm Jupiter origins.

\section{Observations} \label{sec:observation}

\subsection{Summary of Previous Observations}
The planet \toib was detected by the Transiting Exoplanet Survey Satellite \citep[\tess;][]{Ricker14}.
\cite{rodriguez23} first discovered and confirmed its planetary nature using ground-based photometry from the \tess Follow-up Observing Program \citep[TFOP;][]{TFOP}, high-resolution adaptive optics (AO) imaging with the PHARO instrument \citep{PHARO} on the Palomar 200-inch telescope and ShARCS on the Shane 3\,m telescope at Lick Observatory, and high-resolution spectroscopy with the Tillinghast Reflector Echelle Spectrograph (\tres; \href{http://www.sao.arizona.edu/html/FLWO/60/TRES/GABORthesis.pdf}{Fűrész 2008}) on the 1.5\,m Tillinghast Reflector at the Fred Lawrence Whipple Observatory, as well as the MINERVA North telescope array and KiwiSpec Spectrograph \citep{MINERVAN1, MINERVAN2} at Whipple Observatory. The planet's mass and orbit have been constrained.
Later, \cite{chontos24} refined the planet's mass and orbital parameters using the HIRES Spectrograph \citep{HIRES} on the Keck 10-meter telescope on Mauna Kea, Hawaii. 

Here, we briefly summarize the observations used in our modeling.
\changes{The star has been observed in six sectors of \tess—Sectors 25, 26, 40, 52, 53, and 79. As this manuscript was being prepared, the Quick Look Pipeline \citep[QLP;][]{Huang20a, Huang20b} reduced light curves cover Sectors 25, 26, 40, 52, and 53. Notably, the observing cadence decreases from 30 minutes in Sectors 25 and 26 to 10 minutes in Sectors 40, 52, and 53. The Science Processing Operations Center \citep[SPOC;][]{Jenkins16} reduced light curves are available for Sectors 26, 40, 52, 53, and 79, all with an observing cadence of 2 minutes. 
Additionally, 20-second cadence data is available for Sector 79. We use the 2-minute SPOC light curves from Sectors 26, 40, 52, 53, and 79, along with the QLP light curves from Sector 25, for the joint fit.}
Forty HIRES spectra were taken from August 25, 2020, to May 13, 2022, spanning 1.7 years \citep{chontos24}. The median HIRES radial velocity (RV) uncertainty is 5.2\,m\,s$^{-1}$, although this number could be underestimated given the star's large $v\sin{i_\star}$. The HIRES RVs are used for the joint fit with a treatment of the underestimated RV uncertainties.

\subsection{Transit Spectroscopic Observation}
Two transit spectroscopy observations were taken by the NEID spectrograph \citep{NEID_optical, NEID_budget} on the WIYN 3.5\,m telescope at the Kitt Peak National Observatory (KPNO) in Arizona, USA. The NEID spectrograph is a highly stabilized \citep{NEID_performance, NEID_stability}, fiber-fed \citep{NEID_fiber1, NEID_fiber2} spectrograph with a resolving power of $R\approx110,000$ in high-resolution (HR) mode and has wavelength coverage from 380\,nm to 930\,nm.
The first \neid RM-effect visit occurred on May 26, 2023. The observation began at 03:10 \ut and lasted 6.3 hours. We obtained 35 spectra, each with a 10-minute exposure time, in HR mode, covering approximately 69\% of the transit. 
The second \neid visit took place on July 6, 2023. The observation started at 02:30 \ut and lasted 7.5 hours. We obtained 41 spectra, each with a 10-minute exposure time, in HR mode, covering approximately 77\% of the transit.

The \neid data reduction has been performed using three different pipelines: the standard NEID Data Reduction Pipeline v1.3.0 \href{https://neid.ipac.caltech.edu/docs/NEID-DRP/}{(\neiddrp)}, the \serval Pipeline \citep{Zechmeister18, Stefansson22}, and Doppler Tomography \citep[DT;][]{Collier10}. The reduced data are presented in Figure~\ref{fig:rm_visit1} and Figure~\ref{fig:rm_visti2}, respectively.
The \neiddrp pipeline utilizes the cross-correlation function (CCF) technique to extract the radial velocities. Due to a minor bug in the \neiddrp that causes the computed RV error bars to be systematically overestimated for certain targets that have significantly discrepant systemic velocities relative to literature values, we recalculated the RV errors independently using the DRP-derived CCFs using standard techniques \cite{Boisse10}. The median \neiddrp RV uncertainties are 15.4\,m\,s$^{-1}$ and 11.9\,m\,s$^{-1}$ for the first and second visits, respectively.
The \serval pipeline initially builds a stellar template from the NEID observations and uses least-squares fitting to extract the radial velocities. The median RV uncertainties out of the \serval pipeline are 9.5\,m\,s$^{-1}$ and 7.0\,m\,s$^{-1}$ for the first and second visits, respectively. 
Lastly, DT models the line profile variations induced by the transiting shadow of the planet. The line profiles are derived via a least-squares deconvolution \citep[LSD;][]{Donati97} of each observation against a non-rotating synthetic template generated from the ATLAS9 model atmospheres \citep{Castelli:2004}. An average line profile is then removed from each observation, and the residuals are modeled for the planetary transit signature.

\section{Stellar Properties} \label{sec:stellar}
We derive the stellar parameters following the procedures described in Section 4 of \cite{chontos24}. We first use $\mathtt{SpecMatch}$-$\mathtt{Synth}$ \citep{Petigura15} to derive the stellar effective temperature ($T_{\rm eff}$), metallicity ([Fe/H]), and surface gravity ($\log{g}$) of the star. 
We then model the spectral energy distribution (SED) and the MESA Isochrones and Stellar Tracks \citep[MIST;][]{Choi16, Dotter16} to derive the age, mass, and radius of the star using $\mathtt{isoclassify}$ \citep{Huber17}. The stellar parameters $T_{\rm eff}$ and [Fe/H] from $\mathtt{SpecMatch}$ are used as inputs for the model. We include the Johnson $B$ and $V$ magnitudes from the APASS catalog \citep{Henden15}, 2MASS $J$, $H$, and $K_s$ magnitudes \citep{Skrutskie06}, and the \gaia DR3 $G$, $R_p$, and $B_p$ magnitudes \citep{GaiaDR3} to fit the SED. The \gaia DR3 parallax \citep{GaiaDR3} is used to determine the distance to the star. The results are summarized in Table~\ref{tbl:parameters}.
% We also present the stellar evolution track on the HR diagram in Figure~\ref{fig:track}.

\changes{The star has four blocks of \tess data, separated by 1-year or 2-year gaps.} To avoid the dominance of the window function in the periodogram for the entire dataset, we calculate the periodogram of the light curve piece by piece. Interestingly, the Sectors 25--26 \tess data from 2019 show a periodicity of 5.9 days, whereas Sector 40 in 2020 shows 7.0 days, Sectors 52--53 in 2021 show 5.1 days, \changes{and Sector 79 in 2024 shows show 3.4 days. The period detected in Sector 79 is shorter than those in the other sectors, which might indicate that it is an alias of 6.8 days.} We attribute the lack of consistency in the star's periodicity to its multiple spot complexes. The rotation periods between 5--7 days correspond to an equatorial velocity of 20--28\,km\,s$^{-1}$. As a sanity check, this velocity is above the projected rotational velocity of $\sim$18\,km\,s$^{-1}$, and the deviation may indicate a stellar inclination apart from 90 degrees. Although the star is evolved, existing \tess data did not detect the oscillation modes of \toi. According to the scaling relation for the oscillation frequency $\nu_{\text{max}} = 3100\, \mu \mathrm{Hz}\, (M/M_\sun) (R/R_\sun)^{-2} (T_{\mathrm{eff}}/T_{\mathrm{eff}, \sun})^{-0.5}$ \citep{Chaplin19}, \toi should oscillate at $\sim 58$ cycles per day, a signal that is not detected in the \tess data.

% % Stellar evolution tracks
% Going to refit w/ using sky map to see what it estimates for extinction. \textcolor{red}{Doing this now!}

% Stellar evolution \textcolor{red}{Remake tracks with current placement on HR diagram - possibly some other dimensions}

% \begin{figure}
% \centering
% \includegraphics[width=\linewidth]{figures/track.png}
% \caption{Stellar evolution tracks for \toi, assuming a stellar mass of \mstar, an age of \age\ Gyr, and a metallicity of \feh\ [Fe/H].}
% \label{fig:track}
% \end{figure}

\section{Planet Properties} \label{sec:planet}

\begin{figure*}
    \centering
    \includegraphics{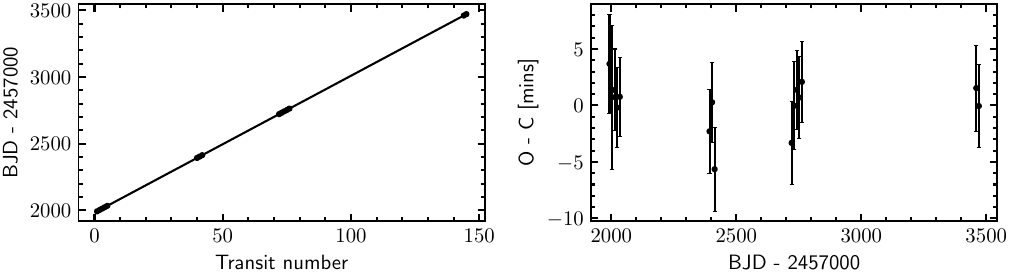}
    \caption{Transit-timing variation on \toib. The planet exhibits a transit timing variation of less than 5 minutes over a \changes{4-year} baseline, with a typical mid-transit time uncertainty of 2.7 minutes. \label{fig:ttv}}
\end{figure*}

\subsection{TTV Modeling}
Since both of our RM-effect measurements captured only a partial transit of \toib, understanding the transit-timing variation (TTV) properties of the planet is crucial for robust stellar obliquity inference. We modeled the \tess transits, treating each transit's mid-transit time as a free parameter. We then fitted a linear line to the mid-transit times, deriving the orbital period, one reference transit epoch, and TTV signals.
Benefiting from multiple sectors of \tess observations over \changes{four} years, we modeled \changes{15} transits of \toib over this period. The results are shown in Figure~\ref{fig:ttv}. The scatter in transit timing variations is less than 5 minutes, with the median mid-transit time uncertainty of about \changes{3.7} minutes. No obvious TTV patterns have been detected in existing \tess observations.
\changes{We derive the orbital period $P = 10.261129 \pm 0.000009$ in days and reference transit epoch $T_C = 1982.49664 \pm 0.00067$ in BJD $-$ 2457000.}

Because of the lack of TTVs, in the global modeling to be discussed in the next section, we model $P$ and $T_C$ without individually modeling each mid-transit time. The derived orbital period and reference transit epoch are well agree with those obtained from the TTV modeling, within 1$\sigma$ consistency.

\begin{figure*}[htb!]
    \gridline{\fig{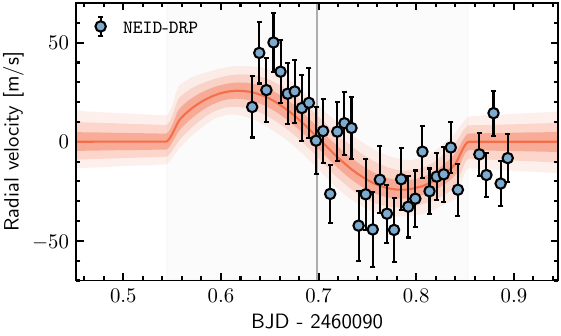}{0.45\textwidth}{\hspace*{1cm}(a) In-transit RVs reduced by the $\mathtt{NEID}$-$\mathtt{DRP\,v1.3.0}$ pipeline, which infer $\lambda=$\DRPlam$\degr$. The orbit trend has been subtracted.}
              \fig{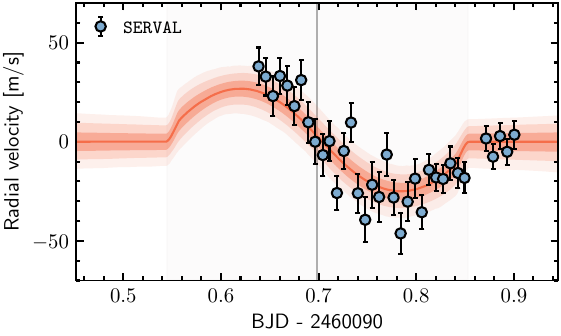}{0.45\textwidth}{\hspace*{1cm}(b) In-transit RVs reduced by the $\mathtt{SERVAL}$ pipeline, which infer $\lambda=$\SERVALlam$\degr$. The orbit trend has been subtracted.}}
    \gridline{\fig{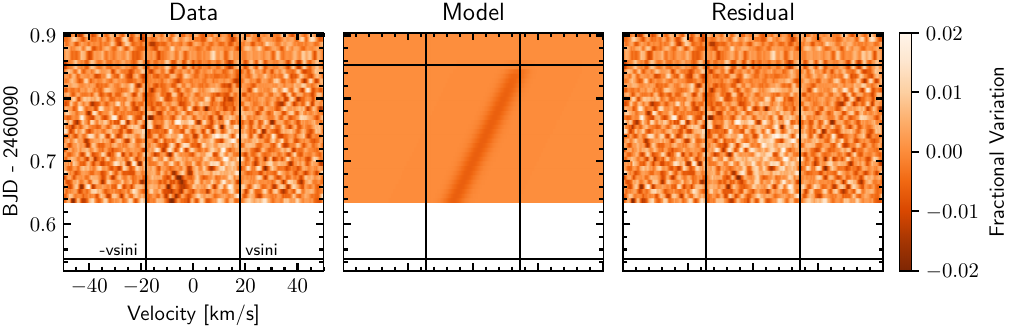}{0.95\textwidth}{(c) Doppler Tomography signal, which infers $\lambda=$\DTlam$\degr$.}}
    \caption{\neid's first RM-effect visit on May 26th, 2023. Data was reduced by three different reduction techniques. The inferred projected stellar obliquities are consistent, but Doppler Tomography (DT) provides the tightest constraint. (a) and (b) In-transit radial-velocity measurements of the \toi system using the \neid spectra. The blue dots and black bars are \neid RVs and their corresponding uncertainties. The planet's transit and mid-transit time are indicated by the grey shaded region and grey vertical line, respectively. (c) The Doppler Tomography (planetary shadow) signal of the \toi system during \toib's transit. The left, middle, and right panels are data extracted from the \neid spectra, best-fit model, and the residual of the data after subtracting the best-fit model. The color scale presents the flux variation of the velocity channel.\label{fig:rm_visit1}}
\end{figure*}

\begin{figure*}[htb!]
    \gridline{\fig{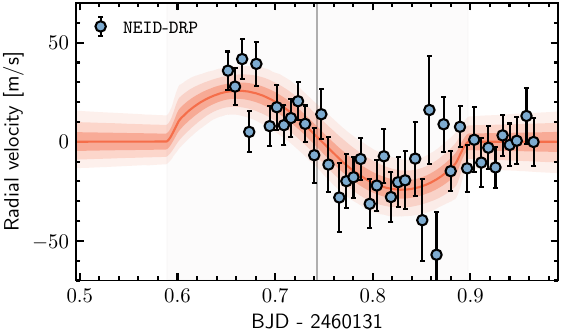}{0.45\textwidth}{\hspace*{1cm}(a) In-transit RVs reduced by the $\mathtt{NEID}$-$\mathtt{DRP\,v1.3.0}$ pipeline, which infer $\lambda=$\DRPlam$\degr$. The orbit trend has been subtracted.}
              \fig{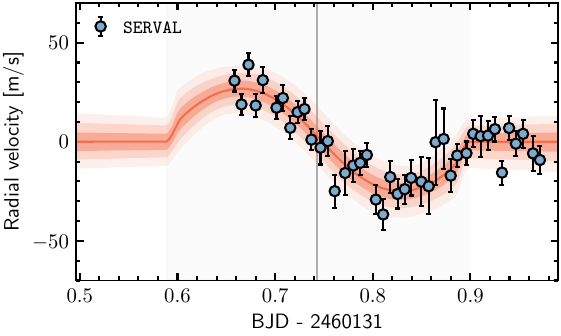}{0.45\textwidth}{\hspace*{1cm}(b) In-transit RVs reduced by the $\mathtt{SERVAL}$ pipeline, which infer $\lambda=$\SERVALlam$\degr$. The orbit trend has been subtracted.}}
    \gridline{\fig{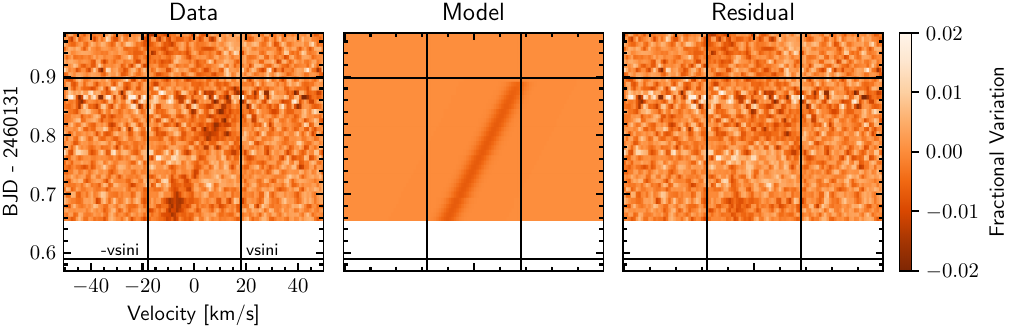}{0.95\textwidth}{(c) Doppler Tomography signal, which infers $\lambda=$\DTlam$\degr$.}}
    \caption{Same as Figure~\ref{fig:rm_visit1}, but for the NEID's second RM-effect visit on July 6th, 2023.\label{fig:rm_visti2}}
\end{figure*}

\subsection{Global Modeling: Transit+RV+RM-effect}
As the main result of this work, we present the joint model \tess transit, \hires radial velocities, and \neid RM-effect signals to derive the planetary and orbital properties of \toib. We use the $\mathtt{exoplanet}$ package \citep{exoplanet:exoplanet, exoplanet:joss} to build the model and perform the Markov chain Monte Carlo (MCMC) using the $\mathtt{PyMC}$ package \citep{pymc2023}.

We run three different models, all including \tess and \hires data, but one with the RM-effect signal reduced by \neiddrp, one by \serval, and one by DT.
The model includes the following planetary and orbital parameters:
\begin{itemize}
    \setlength\itemsep{0em}
    \item $P$: orbital period
    \item $T_C$: reference transit epoch
    \item $b$: impact parameter
    \item $R_p/R_\star$: planet-to-star radius ratio
    \item $M_p$: planet mass
    \item $e$: orbital eccentricity
    \item $\omega$: argument of periapse
    \item $\lambda$: projected stellar obliquity
\end{itemize}
Among these free parameters, a uniform prior is used on $P$, $T_C$, $b$, and $\lambda$, a log-uniform prior is used on $R_p/R_\star$ and $M_p$, and a unit disk vector is used on $\sqrt{e}\cos{\omega}$ and $\sqrt{e}\sin{\omega}$, where both $e$ and $\omega$ are uniformly distributed.
For both the \neiddrp and \serval fittings, we use the \cite{Hirano11} model to calculate the RV anomaly due to the RM effect. In addition to the quadratic limb darkening coefficients \citep{Kipping13} for the \tess transits, we model another pair for the \neid observations. Additionally, we model RV jitters, $\sigma_{\rm RV, DRP}$ and $\sigma_{\rm RV, SERVAL}$, in log-uniform space as free parameters added to both \hires and \neid RV uncertainties.
Lastly, we model the projected stellar rotation velocity $v\sin{i_\star}$ using a Normal prior derived from the spectra.
Independently, we perform the joint modeling of the \tess transit, \hires RV, and \neid DT signals. We model the planetary shadow at each time snapshot as a Gaussian profile that is broadened by the instrumental resolution and macroturbulence of the host star $v_{\rm marco}$, which follows a prior uniformly between 0 and 10\,km\,s$^{-1}$. The center of the velocity profile depends on the projected stellar obliquity $\lambda$ and will be inferred.
For each model, we begin with an optimization and then run the MCMC with 5000 tuning steps and 3000 draws with 4 independent chains. To check the convergence and sampling efficiency, we use the Gelman-Rubin diagnostic \citep[$\hat{\mathcal{R}}$ convergence to 1;][]{Gelman92} and the effective sample size \citep[ESS;][]{Gelman14}. 
All three models have passed the convergence test. A summary of the results can be found in Table~\ref{tbl:parameters} and Figure~\ref{fig:rm_visit1} and \ref{fig:rm_visti2}.

The \neiddrp and \serval pipelines infer $\lambda=$\DRPlam$\degr$ and $\lambda=$\SERVALlam$\degr$, respectively, while the DT signal infers $\lambda=$\DTlam$\degr$. The three inferred projected stellar obliquities are consistent with each other, ruling out a polar or retrograde orbit of \toib. Because of the high dimensionality of the DT signal, the DT model provides the tightest constraint on the projected stellar obliquity $\lambda$. We adopt the DT results for the discussion of this work.

In summary, \toib is a \DTmp\,\mj\,planet on a moderately eccentric ($e_p=$ \DTecc), slightly misaligned ($\lambda=$ \DTlam$\degr$) orbit. The planet's orbital period is $P=$ \DTper-day, with a semi-major axis of $a=$ \DTap-au and a planet-star separation of $a/R_\star=$ \DTaor. The planet has a size of $R_p=$ \DTrp\,\rj. The planet's mass, eccentricity, and radius are consistent with previous estimates \citep{rodriguez23, chontos24}.

In all three models, the \hires RVs present high RV jitters in residuals, with an amplitude of 23\,m\,s$^{-1}$.
We use the Gaussian process kernels for granulation and oscillations given in \cite{Luhn23} to estimate the expected white-noise-equivalent levels of additional variability due to granulation and oscillations. The granulation kernel is composed of two Harvey-like components with frequencies and amplitudes scaled by the effective temperature of \teff\,K and $\log{g}$ of \logg; the expected white-noise equivalent for granulation is 1.6\,m\,s$^{-1}$. The oscillations kernel is described by a stochastically driven, damped harmonic oscillator with frequency and amplitude scaled by $\nu_{max}$; the expected white-noise equivalent for oscillations is 1.7\,m\,s$^{-1}$.
The large RV jitters are likely due to the high $v\sin{i_\star}$ of the host star, which is $\sim$18\,km\,s$^{-1}$.

\subsection{Search for External Companions}
Next, we search for additional planets or stellar companions in the \toi system. Given \toib's high mass and eccentric orbit, \changes{its external perturbers could potentially be massive}. Despite the long baseline of \hires RVs, we find no clear evidence of additional companions in the RV residuals due to the large RV jitters caused by the fast rotation of the host star. The median RV residual is at a level of 12\,m\,s$^{-1}$, while the RV jitter combined with HIRES measurement uncertainty is about 24\,m\,s$^{-1}$.

The proper motion anomaly (PMa) technique is a powerful approach for searching for long-period, massive companions orbiting \toi. This technique is based on identifying the difference between the long-term proper motion of the star, as measured by \gaia and \hippa, and the short-term proper motion recorded by \gaia alone \citep{Brandt21, Kervella22}. The anomaly could indicate the presence of an external companion, though it is subject to degeneracy in mass and semimajor axis. We adopt the PMa for \toi (HIP 86040; Gaia DR3 1344163891352965632) from the \cite{Kervella22} catalog. 
The star exhibits a tangential velocity anomaly of $66.5 \pm 44.8$\,m\,s$^{-1}$. While the signal is insignificant, if the velocity anomaly of the star is indeed introduced by another planet, it could correspond to a $9 \pm 5$ \mj\,-mass planet at 5 au with mass-semimajor axis degeneracy. However, the contributions from stellar noise and instrumental systematics are not well understood, and thus the detection of the companion is inconclusive.
Gaia DR4 may provide more evidence on the existence of external companions.

We also check if \toi has a common proper motion (CPM) with any other stars from the \cite{Kervella22} catalog. While a nearby star with a projected linear separation of 47100 au and a V-band magnitude of 15.9 is found, it receives a candidate companion score ($P_{\rm tot}$) of 0.112 (see Section 3.4.3 of \cite{Kervella22} for a detailed description of the metrics), indicating a low probability of being a co-moving or bound companion. The star is likely a nearby field star.

\section{Discussion}\label{sec:discussion}
\toib is a 10.26-day, $5.7\pm0.3$ Jupiter-mass planet orbiting an evolved A star. The planet has a moderately eccentric orbit of $e = 0.21\pm0.01$. In this work, we combine \tess and previous \hires observations \citep{chontos24} with our new \neid RM-effect measurements to constrain the planet's orbital properties. We find that \toib has a nearly aligned orbit with a projected stellar obliquity of $\lambda=$\DTlam$\degr$.
Given the current low orbital eccentricity of \toib, the planet is unlikely to be undergoing high-eccentricity tidal migration unless it can excite its eccentricity to much higher values, for example, through secular interactions with other planets \citep{Petrovich16}.
\toib more likely migrated inwards from the outer disk or formed in-situ. The observed orbital eccentricity and inclination of \toib is likely an outcome of the planet-disk interactions \citep{Duffell15} or post-formation dynamical evolution with other planets in the system.

The \toi system is interesting from multiple perspectives.
First, \toib is a super Jupiter, a class of giant planets with masses beyond $\sim2$\,\mj\, but still below the brown dwarf's deuterium fusion limit. Recently, \cite{Gupta24} showed that these super Jupiters tend to have more eccentric orbits than their less massive counterparts, potentially as a result of planet-planet interactions. Thus, it is interesting to understand whether \toib is born with this mass or grows due to collisions.
Second, \toib's orbit is nearly aligned with its host star spin axis. We discuss how spin-orbit coupling may play a role in the planet's orbital obliquity.
Third, \toib's host star is evolved. Because of the inflation of the host star radius, its planet-star separation ($a/R_\star$) decreased by a factor of 2 from roughly 18 to 9. The inflation of the star could potentially increase the planet-star tidal interactions, speeding up the spin-orbit alignment or heating the atmosphere of \toib. However, we do not see an inflation of \toib's radius, likely due to the high surface gravity of the planet and also the relatively short timescale since the star evolved off the main sequence.

\subsection{Origins of Super Jupiters}

\begin{figure*}
    \centering
    \includegraphics[width=\linewidth]{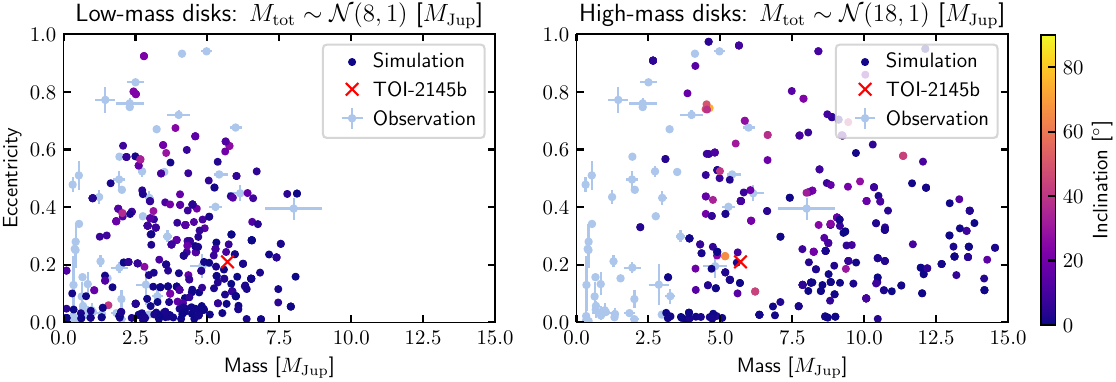}
    \caption{Mass and orbital properties of the innermost warm Jupiter in low-disk mass and high-disk mass scenarios. \toib is labeled as a red cross. Both planets born in a low-mass disk followed by collisions and in a high-mass disk followed by scatterings could explain the current properties of \toib. \label{fig:sims}}
\end{figure*}

\toib is noteworthy for its substantial mass. As a super Jupiter with nearly six times the mass of Jupiter, it raises the question: Was it born with such a high mass, or did the planet acquire its mass through collisions between multiple planets? The observed positive mass-eccentricity relationship among warm, giant planets \citep{Gupta24} may suggest the latter scenario, especially when considering \toib within a population of warm Jupiters. 

Confirming this hypothesis requires a detailed, population-level dynamical study. Here, we build toy models with \emph{N}-body simulations to explore two formation scenarios. We consider a 4-planet system with the innermost planet having a semimajor axis of 0.1 au. 
While our understanding of the multiplicity of giant planets remains largely incomplete, RV surveys, such as the California Legacy Survey \citep[CLS;][]{CLS}, suggest that a significant fraction ($\sim$ 40\%) of giant planets are in multiplanet systems \citep{Zhu22}. Moreover, the 4-planet setup facilitates the formation of super Jupiters through collisions between Jupiter-mass objects.

To construct the initial condition of a planetary system, we use three free parameters: the total mass of the four planets ($M_{\mathrm{tot}}$), the standard deviation of mass distribution among the four planets in the same system ($\sigma_{m_p}$), and the mutual Hill radii between neighboring planet pairs ($\Delta{a_p}$). The masses of the planets in the same system are drawn from a normal distribution with a mean of $M_{\mathrm{tot}}/4$ and a variance of $\sigma_{m_p}$. The initial eccentricity and inclination are assumed to be 0.01 for all planets.
In all population synthesis simulations presented in this work, we assume $\sigma_{m_p}$ follows a normal distribution with a mean of 1 and a standard deviation of 0.5 in the unit of Jupiter mass, bounded between 0 and 2. $\Delta{a_p}$ follows a normal distribution with a mean of 4 and a standard deviation of 0.2.
We vary the $M_{\mathrm{tot}}$ to explore how it determines the outcome of the planetary system architecture.

We consider a low disk mass scenario with a total planet mass $M_{\mathrm{tot}}$ following $\mathcal{N}(8, 1)$ and a high disk mass scenario with $M_{\mathrm{tot}} \sim \mathcal{N}(18, 1)$, both in units of Jupiter mass. We note that the definitions of low and high disk mass here are relative to each other in two scenarios.
We simulate each system for \changes{10} Myr using $\mathtt{REBOUND}$ with the $\mathtt{IAS15}$ integrator \citep{Rein12, Rein15}.
Collisions are checked at each timestep for crossing pathways using the $\mathtt{line}$ algorithm and resolved using the $\mathtt{merge}$ module, which conserves mass, momentum, and volume, but not energy.
The realization of mass, orbital eccentricity, and inclination for the innermost warm Jupiter are shown in Figure~\ref{fig:sims}. \toib is indicated by a red cross. \changes{Observed warm Jupiters are plotted in blue dots for reference.}

\changes{For the low-disk mass case, as shown in the left panel of Figure~\ref{fig:sims}, massive planets above 5 Jupiter masses are mostly grown through collisions, as is the case for \toib. These planets first have their eccentricities excited by interactions with other planets in the system, including scattering and ejection. Later on, collisions happen and tend to reduce the eccentricity of the planets. Planets between 2 and 5 Jupiter masses are also in systems with significant scattering and ejection, and thus experience eccentricity excitation. However, they have fewer mergers than those above 5 Jupiter masses. Planets below 2 Jupiter masses typically have low eccentricities; those with higher eccentricities are often dynamically unstable and get ejected. Collisions with the host star are rare.
For the high-disk mass case, as shown in the right panel of Figure~\ref{fig:sims}, massive planets are more common. For planets with a mass similar to \toib, scattering and ejection are still the dominant dynamical mechanisms to excite their eccentricities. Collisions, however, happen less frequently to these planets in the high-mass disk case than in the low-mass disk case. Therefore, super Jupiters born in low-mass disks and formed via collision are expected, on average, to have lower eccentricities than those born massive.}
A similar trend is observed in the inclination distribution, although it is less pronounced.

As shown by our simulations, \toib can be reproduced in both formation scenarios. It is plausible for the planet to form from a low disk mass followed by collisions or from a massive disk with little or no collision. However, the two proposed scenarios for the formation of super Jupiters lead to distinct predictions about the overall eccentricity distribution of super Jupiters: super Jupiters born massive tend to have a broader eccentricity distribution than super Jupiters grown out of collisions. Based on existing observations of warm Jupiters, \changes{shown as blue dots in Figure~\ref{fig:sims}}, the collision scenario could provide a better match to the data.
Additionally, \changes{our simulations show that} super Jupiters formed through collisions are \changes{generally} expected to have companions with masses similar to or lower than that of regular Jupiters, whereas those born with inherently high masses are likely to have companions with comparable masses. Consequently, searching for companion planets could be important in understanding the mass distribution within the system and determining which formation scenario is more plausible. \changes{This trend is based on the assumption of initial mass similarity among the planets, which will need to be examined in the future.}

\subsection{Spin-Orbit Coupling of Close-in Planets}
\toib has joined the group of about two dozen warm Jupiters that have spin-orbit measurements, many of which show a tendency towards spin-orbit alignment around single stars \citep[e.g.,][]{Rice22, Dong22, Espinoza23, Bieryla24, Radzom24, WXY2024}. It is unclear if such a trend persists in binary systems. For example, TOI-1859b is a 64-day warm Jupiter with an eccentric and misaligned orbit ($e = 0.57^{+0.12}_{-0.16}$, $\lambda = 38.9^{+2.8}_{-2.7}\degr$), whose host star has a distant companion \citep{Dong23}. The role of the binary companion in determining planet formation remains open for discussion.
Here we discuss the importance of spin-orbit coupling and how it might affect stellar obliquity distribution. The gravitational coupling between the close-in giant planet and its oblate host star may prevent the spin-orbit misalignment of the giant planet's orbit excited by the companion. 
Under the assumption of the dynamical perturbation of warm Jupiters happen mostly after they migrate at the current orbital distances, the external companion needs to overcome spin-orbit coupling from the star to excite the inner planet's inclination. Such an effect is the strongest around the fast rotating host stars, which could be \toi in this case.
\cite{Lai18} defined the planet-star coupling factor $\epsilon_{\star1}$, where the smaller the value, the stronger coupling between the planet and the star, and the weaker the distant, external perturber to excite the inclination of the planet.
Using Equation~(24) in \cite{Lai18}, the planet-star coupling factor $\epsilon_{\star1}$ follows
\begin{equation}
    \begin{aligned}
        \epsilon_{\star 1} &= \frac{\omega_{12}}{\omega_{\star 1}} 
        \left( \frac{1 - \omega_{\star 2}/\omega_{1 2}}{1 + S_{\star}/L_1} \right) \\
        &\simeq 1.25 \left( \frac{6k_{q\star}}{k_{\star}} \right)^{-1} 
        \frac{m_2}{m_1} \left( \frac{a_1}{0.04 \, \text{au}} \right)^{9/2} 
        \left( \frac{\tilde{a}_2}{1 \, \text{au}} \right)^{-3} 
        \left( \frac{P_{\star}}{30 \, \text{d}} \right) \\
        &\quad \times \left( \frac{M_{\star}}{M_{\odot}} \right)^{1/2} 
        \left( \frac{R_{\star}}{R_{\odot}} \right)^{-3} 
        \left( \frac{1}{1 + S_{\star}/L_1} \right),
    \end{aligned}
\end{equation}
where $S_{\star}/L_1$ is the ratio of stellar spin angular momentum and orbital angular momentum of the inner planet,
\begin{equation}
    \begin{aligned}
        \frac{S_{\star}}{L_1} &= 0.079 \left( \frac{k_{\star}}{0.06} \right)^{-1} 
        \left( \frac{m_1}{M_J} \right)^{-1} 
        \left( \frac{a_1}{0.04 \, \text{au}} \right)^{-1/2} 
        \left( \frac{P_{\star}}{30 \, \text{d}} \right)^{-1} \\
        &\quad \times \left( \frac{M_{\star}}{M_{\odot}} \right)^{1/2} 
        \left( \frac{R_{\star}}{R_{\odot}} \right)^2.
    \end{aligned}
\end{equation}
Here, the notation $\star$ means the central star, $1$ means the inner planet, and $2$ means the distant perturber. $\omega_{xy}$ means the precession rate of $x$ due to $y$.
$k_{q\star}$ and $k_{\star}$ are the Love numbers of the star and the planet, $m_1$ and $m_2$ are the masses of the planet and the perturber, respectively, $a_1$ is the planet's semimajor axis, $\tilde{a}_2$ is the perturber's effective semimajor axis $\tilde{a}_2 = a_2 \sqrt{1-e_2^2}$, $P_{\star}$ is the star's rotation period, and $M_{\star}$ and $R_{\star}$ are the star's mass and radius.

\begin{figure}
    \centering
    \includegraphics{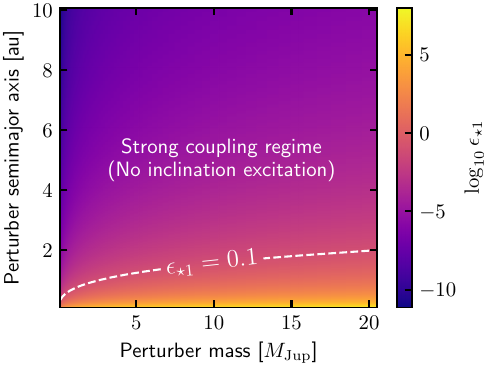}
    \caption{Spin-Orbit coupling factor $\epsilon_{\star 1}$ between \toi and \toib given different perturber properties. Here we assume a stellar rotation rate of 5 days. The dashed line corresponds to $\epsilon_{\star 1} = 0.1$. Perturbers above the line are unlikely to excite the mutual inclination of the inner planet due to spin-orbit coupling.}
    \label{fig:coupling}
\end{figure}

In Figure~\ref{fig:coupling}, we show how the star-planet coupling factor, $\epsilon_{\star 1}$, varies for a perturber with different masses and semimajor axes. If $\epsilon_{\star 1} \ll 1$, we consider the star-planet coupling is strong, and thus it is unlikely that the perturber will excite the inclination of the close-in giant planet. 
As shown in Figure~\ref{fig:coupling}, for a perturber with a semimajor axis greater than $\sim$1--2 au, no matter of its mass, regardless of its mass, the inner planet is always in the strong spin-orbit coupling regime with the star, resulting in a consistently low stellar obliquity.
For example, the hypothetical planet with a semimajor axis of 5 AU and a mass of 9 \mj\, inferred from the \gaia and \hippa proper motion anomalies, would be too distant from TOI-2145b to excite its inclination.
If the perturber is close in ($a_2 \lesssim 1$ au), it may overcome the gravitational coupling. Although this is not indicated by the current stellar obliquity measurements, the existence of such a perturber could potentially be detected in long-term radial-velocity observations. However, the RV precision might be compromised due to the star's high $v\sin{i_\star}$.

\subsection{Stellar Obliquity of Evolved Stars}
The RM-effect observation around evolved stars is challenging due to the increased transit duration caused by stellar radius inflation. Notably, \toib has the longest orbital period among planets orbiting evolved stars for which an RM-effect measurement has been obtained.
It joins a small population of planets, including WASP-71b \citep{Smith13, Brown17}, HAT-P-7b \citep{Winn09, Narita09, Albrecht12, Lund14}, TOI-1181b \citep{Saunders24}, TOI-4379b \citep{Saunders24}, and TOI-6029b \citep{Saunders24}. These planets have orbital periods ranging from 2--6 days and masses of a few Jupiter masses. Most of these planets, except HAT-P-7b, indicate a low stellar obliquity.

\toi is a hot star that had an effective temperature near or above the Kraft break before becoming a subgiant. While stellar evolution may decrease the planet-star separation ($a/R_\star$), thereby increasing star-planet interactions and speeding up the tidal realignment process, whether the planet had a spin-orbit misalignment before stellar evolution remains open for discussion.
As discussed in the previous subsection, spin-orbit coupling between the star and the planet may prevent the excitation of \toib's mutual inclination relative to the perturber planet in the first place.

\begin{table*}
  \centering
  \tabletypesize{\small}
  \caption{Median values and 68\% highest density intervals (HDI) for the stellar and planetary parameters of the \toi (\tic) system. The planetary and orbital parameters are derived from a joint fit of TESS transits, HIRES, and NEID radial velocities. \label{tbl:parameters}}
  \begin{tabular}{llcccc}
    \hline
    \hline
  Parameter & Units & Values \\
  \hline\\\multicolumn{2}{l}{Stellar Properties}&\smallskip\\
  ~~~~$\alpha_{\rm J2016}$\dotfill & \gaia DR3 RA (HH:MM:SS.ss)\dotfill & \ra\\
  ~~~~$\delta_{\rm J2016}$\dotfill & \gaia DR3 Dec (DD:MM:SS.ss)\dotfill & \dec\\
  ~~~~$\varpi$\dotfill & \gaia DR3 parallax (mas) & \parallax\\
  ~~~~$G$\dotfill & \gaia DR3 $G$ magnitude \dotfill & \Gmag\\
  ~~~~$G_{\mathrm{BP}}$\dotfill & \gaia DR3 $G_{\mathrm{BP}}$ magnitude \dotfill & \GBPmag\\
  ~~~~$G_{\mathrm{RP}}$\dotfill & \gaia DR3 $G_{\mathrm{RP}}$ magnitude \dotfill & \GRPmag\\
  ~~~~$M_\star$\dotfill & Stellar mass (\msun)\dotfill & \mstar\\
  ~~~~$R_\star$\dotfill & Stellar radius (\rsun)\dotfill & \rstar\\
  ~~~~$\rho_\star$\dotfill & Stellar density ($\rho_\sun$)\dotfill & \rhostar\\
  ~~~~$\log{g}$\dotfill & Stellar surface gravity (cgs)\dotfill & \logg\\
  ~~~~$T_{\rm eff}$\dotfill & Stellar effective temperature (K)\dotfill & \teff\\
  ~~~~$[{\rm m/H}]$\dotfill & Stellar bulk metallicity (dex)\dotfill & \feh\\
  ~~~~Age\dotfill & Stellar age (Gyr)\dotfill & \age\\
  ~~~~$v\sin{i_\star}_{\rm , spec}$\dotfill & Spectral projected line broadening ($\mathrm{km\,s}^{-1}$)\dotfill & \vsini\\
  \\
  \hline\\\multicolumn{2}{l}{Planetary and Orbital Properties}\smallskip\\
  && With DT & With $\mathtt{SERVAL}$ & With $\mathtt{NEID}$-$\mathtt{DRP}$\smallskip\\
  ~~~~$P$\dotfill & Period (days)\dotfill & \DTper & \SERVALper & \DRPper\\
  ~~~~$T_C$\dotfill & Mid-transit time (BJD-2457000)\dotfill & \DTmidt & \SERVALmidt & \DRPmidt\\
  ~~~~$a$\dotfill & Semi-major axis (au)\dotfill & \DTap & \SERVALap & \DRPap\\
  ~~~~$a/R_\star$\dotfill & Planet-star separation\dotfill & \DTaor & \SERVALaor & \DRPaor\\
  ~~~~$b$\dotfill & Impact parameter \dotfill & \DTb & \SERVALb & \DRPb\\
  ~~~~$i$\dotfill & Orbital inclination ($\degr$)\dotfill & \DTinc & \SERVALinc & \DRPinc\\
  ~~~~$R_p/R_\star$\dotfill & Planet-star radius ratio\dotfill & \DTrprs & \SERVALrprs & \DRPrprs\\
  ~~~~$R_p$\dotfill & Planet radius (\rj)\dotfill & \DTrp & \SERVALrp & \DRPrp\\
  ~~~~$M_p$\dotfill & Planet mass (\mj)\dotfill & \DTmp & \SERVALmp & \DRPmp\\
  ~~~~$e$\dotfill & Orbital eccentricity\dotfill & \DTecc & \SERVALecc & \DRPecc\\
  ~~~~$\omega$\dotfill & Argument of periapse ($\degr$)\dotfill & \DTomega & \SERVALomega & \DRPomega\\
  ~~~~$\lambda$\dotfill & Projected stellar obliquity ($\degr$)\dotfill & \DTlam & \SERVALlam & \DRPlam\\
  \\
  \hline\\\multicolumn{2}{l}{Other Parameters in the Joint Model}\smallskip\\
  ~~~~$v\sin i_\star$\dotfill & Fitted projected line broadening ($\mathrm{km\,s}^{-1}$)\dotfill & \DTvsini & \SERVALvsini & \DRPvsini\\
  ~~~~$\sigma_{\rm RV, HIRES}$\dotfill & HIRES RV jitter ($\mathrm{m\,s}^{-1}$)\dotfill & \DThiresjitter & \SERVALhiresjitter & \DRPhiresjitter\\
  ~~~~$v_{\rm macro, v1}$\dotfill & Host star macroturbulence, visit 1 ($\mathrm{km\,s}^{-1}$)\dotfill & \nonrotv & - & -\\
  ~~~~$v_{\rm macro, v2}$\dotfill & Host star macroturbulence, visit 2 ($\mathrm{km\,s}^{-1}$)\dotfill & \nonrotvv & - & -\\
  ~~~~$\sigma_{\rm RV, SERVAL, v1}$\dotfill & $\mathtt{SERVAL}$ RV jitter, visit 1 ($\mathrm{m\,s}^{-1}$)\dotfill & - & \servaljitter & -\\
  ~~~~$\sigma_{\rm RV, SERVAL, v2}$\dotfill & $\mathtt{SERVAL}$ RV jitter, visit 2 ($\mathrm{m\,s}^{-1}$)\dotfill & - & \servaljitterr & -\\
  ~~~~$\sigma_{\rm RV, DRP, v1}$\dotfill & $\mathtt{NEID}$-$\mathtt{DRP}$ RV jitter, visit 1 ($\mathrm{m\,s}^{-1}$)\dotfill & - & - & \neidjitter\\
  ~~~~$\sigma_{\rm RV, DRP, v2}$\dotfill & $\mathtt{NEID}$-$\mathtt{DRP}$ RV jitter, visit 2 ($\mathrm{m\,s}^{-1}$)\dotfill & - & - & \neidjitterr\\

  \smallskip\\
  \hline
  \end{tabular}
  \tablecomments{\gaia magnitudes and spectral line broadening parameter are obtained from the \gaia Data Release 3 \citep{GaiaDR3}. Both NEID RM-effect observations are included in the joint fit. Planetary parameters inferred from the Doppler Tomography signal are used for discussion.}
  \end{table*}

\section*{Acknowledgments}
\changes{We appreciate the referee for a thoughtful and detailed report, which helped us improve our paper.}
We thank Zhao Guo for the insightful discussions about the detectability of asteroseismology signals. We appreciate fruitful discussion with Phil Armitage on the origins of super Jupiters and sub-Saturns. We would also like to thank Dong Lai for the insightful discussions about the spin-orbit coupling of close-in planets. Special thanks to Te Han for the development and discussions about the NEID $\mathtt{SpecMatch}$. We extend our gratitude to the Flatiron CCA and the NYC astronomical community, as well as to Nora Eisner, Lehman Garrison, Isabel Colman, Lily Zhao, Quang Tran, and Julianne Dalcanton, for their support in the application of the Lomb-Scargle periodogram for stellar rotation periodicity detections. The Flatiron Institute is a division of the Simons foundation. S.W. acknowledges support from Heising-Simons Foundation grant $\#$2023-4050 and support from the NASA Exoplanets Research Program NNH23ZDA001N-XRP (grant $\#$80NSSC24K0153). This research was carried out, in part, at the Jet Propulsion Laboratory and the California Institute of Technology under a contract with the National Aeronautics and Space Administration. The Center for Exoplanets and Habitable Worlds is supported by the Pennsylvania State University and the Eberly College of Science.

\changes{
This work includes data collected by the TESS mission, which are publicly available from the Mikulski Archive for Space Telescopes (MAST). Funding for the TESS mission is provided by the NASA Science Mission directorate. We acknowledge the use of public TESS data from pipelines at the TESS Science Office and at the TESS Science Processing Operations Center. Resources supporting this work were provided by the NASA High-End Computing (HEC) Program through the NASA Advanced Supercomputing (NAS) Division at Ames Research Center for the production of the SPOC data products.
The TESS data presented in this paper were obtained from the MAST at the Space Telescope Science Institute. The specific observations analyzed can be accessed via \dataset[10.17909/t9-nmc8-f686]{https://doi.org/10.17909/t9-nmc8-f686} and \dataset[10.17909/t9-st5g-3177]{https://doi.org/10.17909/t9-st5g-3177}. The TESS Input Catalog and Candidate Target List can be accessed via \dataset[10.17909/fwdt-2x66]{https://doi.org/10.17909/fwdt-2x66}.
This work has made use of data from the European Space Agency (ESA) mission {\it Gaia} (\url{https://www.cosmos.esa.int/gaia}), processed by the {\it Gaia} Data Processing and Analysis Consortium (DPAC, \url{https://www.cosmos.esa.int/web/gaia/dpac/consortium}).}

Data presented were obtained by the NEID spectrograph built by Penn State University and operated at the WIYN Observatory by NOIRLab, under the NN-EXPLORE partnership of the National Aeronautics and Space Administration and the National Science Foundation. These results are based on observations obtained with NEID on the WIYN 3.5m Telescope at Kitt Peak National Observatory (co-PIs: Ashley Chontos \& Jiayin Dong, NOIRLab 2023A-652300). WIYN is a joint facility of the University of Wisconsin–Madison, Indiana University, NSF's NOIRLab, the Pennsylvania State University, Purdue University, University of California, Irvine, and the University of Missouri. The authors are honored to be permitted to conduct astronomical research on Iolkam Du'ag (Kitt Peak), a mountain with particular significance to the Tohono O'odham.

This research made use of $\mathtt{exoplanet}$ \citep{exoplanet:exoplanet, exoplanet:joss} and its dependencies \citep{exoplanet:agol20, exoplanet:astropy13, exoplanet:astropy18, exoplanet:exoplanet, exoplanet:foremanmackey17, exoplanet:foremanmackey18, exoplanet:kipping13, exoplanet:luger19, pymc}.

\vspace{5mm}

\facilities{TESS, \emph{Gaia}, WIYN/NEID, Keck/HIRES, Exoplanet Archive}

\software{$\mathtt{ArviZ}$ \citep{arviz_2019}, $\mathtt{astropy}$ \citep{exoplanet:astropy13, exoplanet:astropy18}, $\mathtt{celerite2}$ \citep{exoplanet:foremanmackey17, exoplanet:foremanmackey18}, $\mathtt{exoplanet}$ \citep{exoplanet:joss, exoplanet:exoplanet}, $\mathtt{Jupyter}$ \citep{Jupyter}, $\mathtt{Matplotlib}$ \citep{Matplotlib07, Matplotlib16}, $\mathtt{NumPy}$ \citep{NumPy11, NumPy20}, $\mathtt{pandas}$ \citep{mckinney-proc-scipy-2010, reback2020pandas}, $\mathtt{PyMC}$ \citep{pymc}, $\mathtt{SciPy}$ \citep{2020SciPy-NMeth}, $\mathtt{Tapir}$ \citep{Jensen13}}

\bibliography{rm2145}{}
\bibliographystyle{aasjournal}

\end{document}